\documentclass[showpacs,aps,prb,twocolumn,floatfix]{revtex4-1}
\usepackage{bm}
\usepackage[dvipdfm]{hyperref}
\usepackage{graphicx}
\usepackage{epsfig}
\usepackage{amsmath,graphics,epsfig,color,verbatim,ulem}
\usepackage{bbold}
\usepackage{float} 
\makeatletter
\newcommand*{\rom}[1]{\expandafter\@slowromancap\romannumeral #1@}

\begin{document}
\title{Density Functional plus Dynamical Mean-Field Theory of the Spin-Crossover Molecule Fe(phen)$_2$(NCS)$_2$ }
\author{Jia Chen$^{1,2}$, Andrew J. Millis$^2$, Chris A. Marianetti$^1$ }
\address{$^1$Department of Applied Physics and Applied Mathematics, Columbia University, New York, New York 10027, USA}
\address{$^2$ Department of Physics, Columbia University, New York, New York 10027, USA}
\date{\today}

\begin{abstract}
We study the spin-crossover molecule Fe(phen)$_2$(NCS)$_2$ using density
functional theory (DFT) plus dynamical mean-field theory, which allows access
to observables not attainable with traditional quantum chemical or electronic
structure methods. The temperature dependent magnetic susceptibility, electron
addition and removal spectra,  and total energies are calculated and compared
to experiment. We demonstrate that the proper quantitative energy difference
between the high-spin and low-spin state, as well as reasonably accurate values
of the magnetic susceptibility can be obtained when using realistic interaction
parameters. Comparisons to DFT and DFT+U
calculations demonstrate that dynamical correlations are critical to  the
energetics of the low-spin state. Additionally, we elucidate the differences
between DFT+U and spin density functional theory (SDFT) plus U methodologies, demonstrating that DFT+U can recover SDFT+U results
for an appropriately chosen on-site exchange interaction. 
\end{abstract}

\maketitle

The combination of density functional theory (DFT) and dynamical mean-field theory (DMFT) is now established in condensed matter physics as a successful theory of materials with strong local electron correlations.  \cite{RevModPhys.68.13,RevModPhys.78.865} Initially devised as a theory of extended (infinite) systems, the method has been extended to finite systems \cite{Florens07,Valli10,Jakob10,Turkowski10,Lin2011,Dominika2011,Kabir13} and has been used to demonstrate that  many-body effects are important for  ligand binding on the active center of protein myoglobin and haemoglobin. \cite{Weber09042014, Weber2013} As compared to traditional highly accurate quantum chemical methods 
DFT+DMFT provides many advantages including excited-state properties, nonzero temperatures and treatment of arbitrary strength of local correlations. Also, because  the computational cost scales linearly with the number of symmetry-inequivalent correlated atoms, the method can be used to treat molecules containing many transition metal or actinide atoms.  However, its broad applicability and quantitative effectiveness in the quantum chemical context is not yet fully established.

Here we apply the DFT+DMFT method to study  spin-crossover complexes: molecular species that change spin state upon  increase of temperature or other changes in environment. Spin crossover molecules provide an important challenge to theory, requiring both accurate energetics and the ability to treat excitations in a situation that (because of the spin) necessarily involves strong electron correlations. Insights gained from study of spin crossover materials could potentially be useful for the design of  thin films  \cite{Shi2009,Zhang2014} or  single-molecule \cite{Miyamachi2012} spintronic devices.  

This paper provides a comprehensive DFT+DMFT description of Fe(phen)$_2$(NCS)$_2$, \cite{Konig1966} a member of an extensively studied and still expanding family of spin-crossover complexes based on Fe(II). \cite{Gutlich1994,Gutlich2000} We compute the magnetic susceptibility, electron addition and removal spectra, and  total energy, finding results in good agreement with experimental data when using realistic interaction parameters. Our analysis enables us to infer that the metal-to-ligand bond length is the control parameter of spin transition. We explore the sensitivity of various observables to the double-counting correction and the on-site interactions $U$ and $J$, demonstrating that Fe(phen)$_2$(NCS)$_2$ is a useful testbed for current and future first-principles methods. Comparison of our results to those obtained with the Hartree approximation (ie. DFT+U\cite{Anisimov1991, Anisimov1993}), demonstrates the importance of dynamical fluctuations in capturing the physics of strong hybridization that is present in the LS state. 

The key feature of Fe(phen)$_2$(NCS)$_2$ is the octahedrally coordinated Fe(II) ion. \cite{Gutlich1994,Gutlich2000} As the temperature is increased above $T^\star=176K$, there is an abrupt increase in magnetic susceptibility which is believed to be related to a change in the electronic configuration from the nominal low-spin (LS) state $t_{2g}^6e_g^0$ with no unpaired electrons to the nominal high spin (HS) state $t_{2g}^4e_g^2$ with 4 unpaired electrons. The average Fe-to-N distance, $d(Fe-N)$, is also longer by about 0.2\AA ~in the  HS than in the LS state. \cite{Gall1990} 
 Experimental measurements estimate that the  energy splitting between LS and HS states is $E_{HS}-E_{LS}\approx 0.13eV$. \cite{Reiher2002} This difference is much greater than the  contribution $k_BT\cdot ln 5\approx 0.025eV$ to the free energy from  the change in  electronic entropy.

Obtaining a LS-HS energy difference of the correct order of magnitude has proven challenging for theory. Spin density functional theory in the generalized gradient approximation (GGA) level overestimates the stability of low-spin state while Hartree-Fock theory incorrectly predicts HS as the ground state.\cite{Reiher2002} These considerations motivated people to consider hybrid functionals,  which interpolate between DFT and HF energies and therefore can be tuned to obtain the desired energy difference. However, the the amount of exact exchange was found to be less than is normally considered reasonable. \cite{Reiher2002,Slimani2014} An extensive study of spin-crossover molecules using modern density functional theory, including meta-GGA and hybrid meta-GGA and double-hybrid functional, shows, generally speaking, relative energies of spin multiplicities are still challenging for DFT methods. \cite{Ye2010} The spin density functional plus U (SDFT+U) \cite{Anisimov1993,Dudarev1998} method was also applied to this system, but again obtaining the correct energy splitting required choosing   $U \approx 2.5eV$, much smaller than is believed to be relevant for Fe.\cite{Lebegue2008,Bucko2012} Diffusion Quantum Monte-Carlo method has also been applied to charged spin-crossover molecules, although direct comparisons to experiments are not currently feasible for charged systems. \cite{Droghetti2012,Droghetti2013} Very recently a detailed quantum chemical calculation based on 
CASPT2 methods reported an energy splitting of 0.17eV, \cite{Rudavskyi2014} indicating the importance of correlations in this system.

\begin{figure}
\includegraphics[width=0.9\columnwidth,clip]{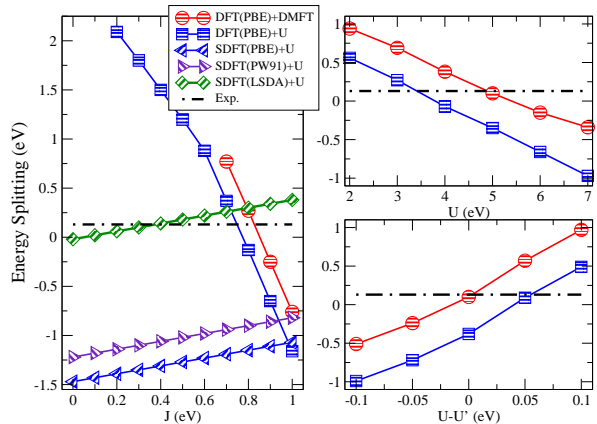}
\caption{Left panel: Energy splitting $E_{HS}-E_{LS}$ as function of J for $U=5eV$. Right upper panel: Energy splitting as function of U for $J=0.85eV$. Right lower panel: Energy splitting as function of  double counting parameter $U^\prime$ (Eq.~\ref{EDC}) for $U=5eV$ and $J=0.85eV$. Temperature is 387K in DMFT calculations. SDFT+U calculations were performed with projector basis in VASP.} \label{LSDA}
\end{figure}

Here we perform fully charge self-consistent DFT+DMFT calculations of the total energy using the method described in Ref.~\onlinecite{Park2014,  Park2014_2}. The DFT part of our calculations use the Vienna ab initio simulation package (VASP), \cite{Kreese1996,Kreese1999} with the Perdue-Burke-Ernzerhof exchange-correlation functional,\cite{Perdew1996} an energy cutoff of 400eV  and a supercell of edge length 15\AA. Maximally-localized Wannier functions (MLWF),\cite{Mostofi2008} constructed using an energy window of 26.5eV with 4.4eV of empty states (120 states total), were used to represent the so-called hybridization window and to construct the correlated subspace\cite{Park2014, Park2014_2} in which DMFT is performed. The hybridization window includes  all 85 occupied states as well as the anti-bonding $\pi$ orbitals on phenanthroline and thiocyanate groups that hybridize with the $t_{2g}$ orbitals of Fe.  The correlated subspace is chosen as  the five $d$-like iron-centered orbitals. Following common practice, for the intra-$d$ interaction we take  the density-density part of the  full on-site Coulomb interaction, parametrized by two independent interaction constants denoted as U and J (we use the form stated in Ref.~\onlinecite{Pavarini2011})  with (unless otherwise specified) U=5.0 eV and J=0.85 eV, consistent with estimates in the literature for Fe in various compounds. \cite{Weber09042014, Weber2013,Kutepov2010,Yin2011}  A double counting term is needed to remove the on-site $d$ interactions present in the Hartree and exchance-correlation functionals. We use the spin-independent form of Anisimov \cite{Anisimov1993} \begin{align}
V_{dc}=U^\prime(N_d-\frac{1}{2})-J(\frac{N_d}{2}-\frac{1}{2})\label{EDC}
\end{align}  
Park et al,  \cite{Park2014, Park2014_2} noted that one should allow for the possibility that the coefficient $U^\prime$ in Eq.~\ref{EDC}  differs from the coefficient $U$ in the interaction. However in this paper we set $U=U^\prime$ everywhere except in the discussion of Fig.~\ref{LSDA}.

The impurity model is solved using the  Continuous-Time Quantum Monte Carlo (CTQMC) in its hybridization expansion  (CT-HYB) form \cite{Werner2006,Gull11_review} implemented by Gull et al \cite{Hafermann2013} in the ALPS package.\cite{Bauer2011} We also solved the impurity problem using a Hartree approximation to the interaction. This is the DFT+U approximation, but implemented using the same correlated subspace and double counting as in the DFT+DMFT calculation, enabling an unambiguous comparison of the results obtained from the two methods.  In both cases the  whole DFT+DMFT loop is iterated until the total energy difference between consecutive updates of charge density is less than 5 meV. Spectral functions are obtained via analytic continuation of the computed imaginary time Green's function using the  maximum entropy method \cite{Jarrell1996}  and the Fe contribution to the  magnetic susceptibility is calculated from the impurity model  spin-spin correlation functions, which are measured in our CTQMC calculations, as 
\begin{align}
\chi_{loc}=\left(g\mu_B\right)^2\int_0^{\beta} d\tau \langle S_z(0)S_z(\tau) \rangle
\label{chiloc}
\end{align}
where $S_z$ is the z component of  spin on the impurity site.

The total energy (see Refs.~\onlinecite{Park2014, Park2014_2}  for details) is 
\begin{multline}
E^{tot}[\rho,\hat{G}_{loc}]=E^{DFT}[\rho]+E^{KS}[\rho,\hat{G}_{loc}]+E^{pot}[\hat{G}_{loc}]-E^{dc}
\label{energy}
\end{multline}
Here $E^{DFT}$ is the density functional theory approximation to the total energy, $E^{KS}[\rho,\hat{G}_{loc}]$ is a correction to the DFT energy arising from the difference between the DFT and DFT+DMFT density matrices, and the interaction energy term  $E^{pot}[\hat{G}_{loc}]$ is calculated from the  frequency dependent self-energy $\hat{\Sigma}$ and Green's function as 
\begin{align}
E^{pot}=\frac{1}{2}T\sum_{n}Tr\left[\hat{\Sigma}(i\omega_{n})\hat{G}_{loc}(i\omega_{n})\right]
\end{align}

\begin{figure}
\includegraphics[width=0.85\columnwidth,clip]{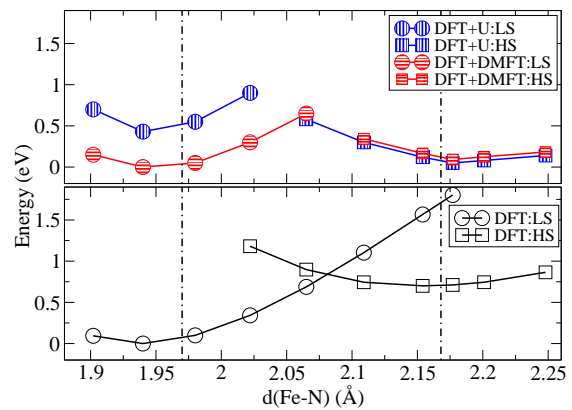}
\caption{Dependence of total energy  on average Fe-to-N bond length. The  DFT+U and DFT+DMFT calculations used  $U=U^\prime=5.0eV$ and $J=0.85eV$ and the DFT+DMFT calculations were done at 387K. The two dashed lines indicate the experimentally measured average Fe-to-N bond length for LS and HS states.\cite{Legrand2007}} \label{tot_energy}
\end{figure}

To perform structural relaxations within our many-body DFT+DMFT theory, we first define reference structures using structurally relaxed DFT calculations. The metastable HS (LS)  state was obtained by using an initialization of the Fe magnetic moment at the nominal high-spin (low-spin) value. We then construct a path between the two structures by linearly interpolating all atomic positions between the values found for the LS and HS structures and minimize the DFT+DMFT energy along this path. We parametrize the path in structure space by the Fe-N bond length.

Fig.~\ref{tot_energy} shows the structure dependence of the  DFT+DMFT energies along with those obtained by density functional and DFT+U methods. All three methods yield two locally stable structures, one with a shorter $d(Fe-N)$, which will be seen to correspond to the LS state, and one with a longer $d(Fe-N)$, which will be seen to be the HS state.  

\begin{table}
\caption{Energy splitting and optimized average Fe-to-N bond length from DFT, DFT+U and DFT+DMFT calculations.}\label{table}
\begin{tabular}{lcccc}
 \hline
Method & DFT & DFT+U & DFT+DMFT&~Expt\cite{Legrand2007,Reiher2002}\\
 \hline
$E_{HS}-E_{LS}$ (eV) & 0.70 & -0.38 &0.10&0.13\\
LS $d(Fe-N)$ \AA  &1.94 &1.94 &1.94&1.97\\
HS $d(Fe-N)$ \AA & 2.15 & 2.20 & 2.18&2.17\\
 \hline
\end{tabular}
\end{table}

\begin{figure}
\includegraphics[width=0.80\columnwidth,clip]{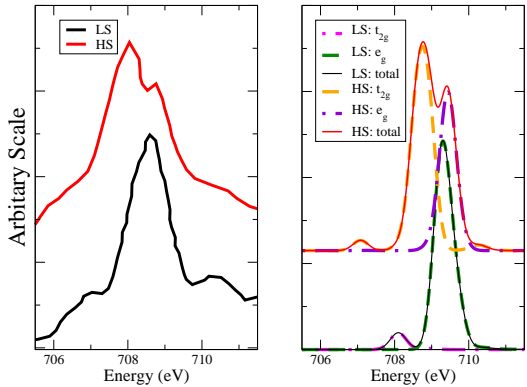}
\caption{(a) Fe L$_{\rom{3}}$ edge X-ray absorption spectroscopy of LS and HS measured at 17K and 298K respectively from Ref.~\citenum{Lee2000}. (b) Spectral functions of empty Fe {\it d} state from analytic continuation via maximum entropy method. Absolute positions of spectral functions are shifted to match experiment.} \label{xray}
\end{figure}

The bond lengths and LS-HS energy splittings computed for the locally stable structures are given in Table ~\ref{table}. While all methods give bond lengths in reasonable agreement with experiment, both DFT and DFT+U methods give an inadequate account of the energy differences between the LS and HS structures. DFT predicts that the LS state is  much too stable while DFT+U, with a physically reasonable U and J, incorrectly predicts the HS state to be the ground state. Similar to hybrid functional calculations \cite{Reiher2002} and previous spin density functional +U calculations,\cite{Lebegue2008, Bucko2012} it is possible to tune U to a value that reproduces the observed energy difference within DFT+U, but the required $U\approx2.5$ eV is unphysically small. By contrast, the DFT+DMFT calculations produce a result in good agreement with experiment with physically reasonable  interaction parameters.

The respective electronic states of the short-bond and long-bond structures are found to be locally stable up to the highest temperatures studied $\sim 1200K$; suggesting, in agreement with deductions from experimental data,\citep{Sorai1974555} that the electronic entropy and energetics are not enough to drive the observed transition. We infer from these results that the Fe-to-N bond length is the critical variable; indeed calculations (see Fig. S2 in Supplemental Material) show that the LS to HS transition occurs when $d(Fe-N)$ crosses a critical value approximately 2.10\AA. Phonon free energy will determine the actual transition temperature. 

\begin{figure}
\includegraphics[width=0.9\columnwidth,clip ]{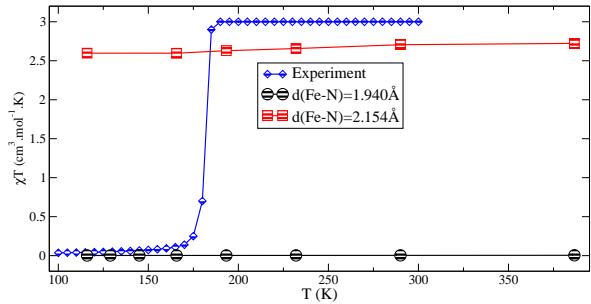}
\caption{Temperature dependence of product $\chi$T of susceptibility (computed from Eq.~\ref{chiloc} with $g=2$) and temperature  from DFT+DMFT calculations for high and low spin structures at $U=5.0eV$ and $J=0.85eV$ compared to  experimental data.\cite{Shi2009}} 
\label{exp}
\end{figure}

From our DFT+DMFT calculations we obtain the many-body density matrix describing the probability of different configurations of the d-orbitals (see Supplemental Material). We find, as expected, that the dominant configuration in the  LS state has zero total spin and  is described by an almost complete occupancy of the $t_{2g}$ symmetry d-states. The key issue for the energetics of the LS state is the correct treatment of the virtual charge fluctuations into the $e_g$ states, in light of the strong Coulomb repulsion associated with multiple occupancy of the $e_g$ states.  DFT  predicts  an $e_g$ occupancy of 1.25 electrons and a relatively large hybridization energy gain. Both the $e_g$ occupancy and the hybridization energy gain are likely excessive due to the inadequate treatment of correlations in current DFT implementations. Alternatively, DFT+U, which adds an extra Hartree term, overestimates the correlation energy, providing an $e_g$ occpuancy of 0.99 electrons (likely too small) and an underestimate of the hybridization energy.   DMFT treats the hybridization more correctly, giving   an $e_g$ occupancy of 1.15 and a reasonable value for the energy of the LS state. The improved properties of DFT+DMFT result from a proper characterization of the multiconfigurational character of LS state,  as also found in quantum chemistry calculations.\cite{Domingo2010} Turning now to the high spin (HS) state, we find that the  HS state is found to be  the $d^5$ maximal spin configuration, with only  small quantum fluctuations towards $d^6$ and lower total spin, leading to a mean d-occupancy $\sim 5.3$. By neglecting the correlated nature of the virtual hopping into the $d^6$ configuration DFT+U allows all of the virtual hoppings to add in parallel, thus  overestimating the hybridization energy gain, but because the mulitconfigurational character of the HS state is weak, the difference between DFT+U and DFT+DMFT is slight.

Having established accurate energetics, we now turn to spectra and magnetic response. X-ray absorption spectroscopy experiments in which incident X-rays are tuned to the Fe L$_{\rom{3}}$ edge probe the empty d-states, revealing information about the electronic configuration of the Fe ions.  Representative data \cite{Lee2000} are compared to our DFT+DMFT calculations in Fig.~\ref{xray}. The theoretical calculations (right panel) reveal that in the LS configuration the density of empty $t_{2g}$ states is very low; the $e_g$ spectrum reveals a single main peak, with a small prepeak of $t_{2g}$ origin arising from a small probability of a LS $d^5$. Alternatively, in the HS situation the two peaks correspond to transitions into the empty $t_{2g}$ and $e_g$ states respectively.  The calculated spectra (for example the $t_{2g}-e_g$ crystal field splitting in the HS state) are in reasonably good agreement with the data although not all of the detailed structure away from the main peaks is reproduced (see Ref.~\onlinecite{Lee2000} for a possible interpretation).

Fig.~\ref{exp} shows the  susceptibility calculated from Eq.~\ref{chiloc} along with experimental results from Ref.~\onlinecite{Shi2009}. We see that the structure with long mean Fe-N bond length indeed has a Curie susceptibility $\chi\sim 1/T$  permitting its identification as the HS state, while the short bond (LS) state has a very small susceptibility. Experimental values are 0.5$\mu_B$ and 5$\mu_B$ for LS and HS, respectively, \cite{Shi2009}    whereas we obtained around 0.004$\mu_B$ for LS and 4.6$\sim$4.8$\mu_B$ for HS. Our calculations only include the d-electron contribution to $\chi$; the nearly quantitative agreement with experiment suggests this is the dominant contribution. 

Crucial to the DFT+DMFT formalism are the values of the  interactions in the correlated subspace and the double counting correction. We have demonstrated  a respectable degree of accuracy using the standard double-counting approach and accepted values for the on-site interactions.  However, it is critical to understand the sensitivity of the results to these approximations. The left panel of Fig.~\ref{LSDA} shows that the DFT+DMFT and DFT+U results for the LS-HS energy difference depend strongly on $J$, as expected since $J$ is the term in the energy favoring locally high spin configurations. The magnitude of the slope is sufficiently large that a relatively small increase in the exchange from $J=0.85eV$ to $J=0.9eV$ would change the sign of the energy splitting, demonstrating the importance of precisely knowing the exchange. However it is significant that that generally accepted value $J=0.85eV$ yields an exchange splitting with the correct $\sim 0.1eV$ order of magnitude.

Fig.~\ref{LSDA} also shows that three widely used SDFT+U methodologies yield a qualitatively different result as compared to DFT+U (and DFT+DMFT) in two key respects: the magnitude and sign of the slope and the value of the $J=0$ intercept.  The difference in $J=0$ values of the energy splitting shows that the SDFT functionals have an effective J built in to them; comparison to the J-dependent DFT+U results demonstrates that  $J_{eff}$ for SDFT(LSDA)+U, SDFT(PW91)+U, and SDFT(PBE)+U are approximately $J_{eff}\approx 0.75eV$, $J_{eff} \approx 0.93eV$, and $J_{eff}\approx 1.05eV$, respectively.  The counterintuitive finding that the SDFT+U energy splitting increases with increasing $J$  can be traced to the spin-dependent double counting correction, which overcompensates the effect of the $J$ in the interaction,  increasing the energy splitting.  This is reasonable behavior given that all of the spin-dependent exchange correlation functionals incorrectly predict the energy splitting to be negative at $J=0$.  Therefore, careful analysis is required  in the use of SDFT theories as a base on which to build a correlated calculation.

The right upper panel of Fig.~\ref{LSDA} presents the U dependence of the HS-LS energy splitting. We see that for each method, a $U$  can be found that reproduces the measured energy difference, and the trends with U are similar in all methods, but the DFT+DMFT method gives the physically correct splitting when a reasonable $U$ is employed.  Motivated by previous work on rare earth nickelates, \cite{Park2014, Park2014_2} we show in the right lower panel the effects of varying the double counting correction, setting $U^\prime\neq U$ in Eq.~\ref{EDC}.  For given $U$, $J$ and $U^\prime$ , the DFT+DMFT procedure always yields a smaller energy difference than the DFT+U methodology.  The dependences illustrated in Fig.~\ref{LSDA} indicate that in order for the method to become truly predictive, improved theoretical understanding of the interaction parameters and double counting is required. Results presented here can serve as benchmarks for this endeavor. 

In summary, we have shown that DFT+DMFT with  generally accepted interaction parameters produces energetics, magnetic susceptibilities, and x-ray absorption spectra in reasonable agreement with experimental measurements on Fe(phen)$_2$(NCS)$_2$.  The method involves a  full self-consistency between the correlated subspace and the background, but the locality assumption basic to many solid-state applications of DMFT is here exact because there is only one correlated site. The ability of DFT+DMFT to handle hybridization in a correlated environment is important for the success of the method. The ability to perform  calculations for a range of temperatures and structures revealed that electronic energy and entropy considerations do not account for the observed transition. The mean  metal-to-ligand bond length is the key parameter controlling the spin state and the transition is likely driven by phonon free energy considerations. 

The authors would like to thank Hyowon Park for very helpful discussions. This research is supported by U.S. Department of Energy under Grant No. DE-SC0006613. AJM also acknowledges DE-ER046169. We used resources of the National Energy Research Scientific Computing Center, a DOE Office of Science User Facility supported by the Office of Science of the U.S. Department of Energy under Contract No. DE-AC02-05CH11231.


\bibliography{Collection}

\end{document}


\title{Supplementary Material for\\ Density Functional plus Dynamical Mean-Field Theory of Spin-Crossover Molecule Fe(phen)$_2$(NCS)$_2$}

\author{Jia Chen$^{1,2}$, Andrew J. Millis$^2$, Chris A. Marianetti$^1$ }
\address{$^1$Department of Applied Physics and Applied Mathematics, Columbia University, New York, New York 10027, USA}
\address{$^2$ Department of Physics, Columbia University, New York, New York 10027, USA}

\maketitle




\section{Magnetic susceptibility}

\begin{figure}[htb]
\includegraphics[width=\linewidth,clip]{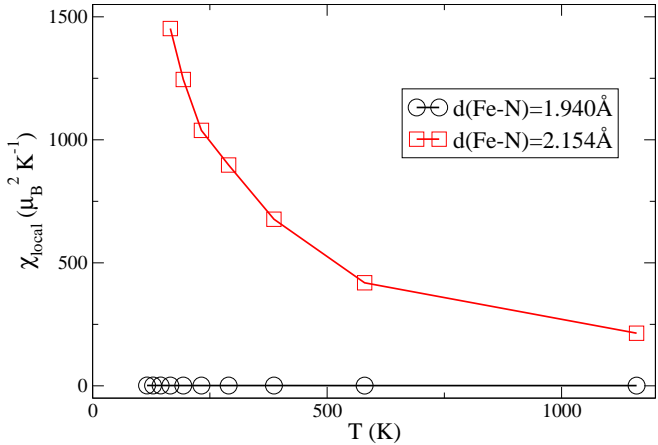}
\caption{Temperature dependence of the Fe magnetic susceptibility $\chi$ from DFT+DMFT calculations in LS and HS structures with $U=5eV$ and $J=0.85eV$.} \label{curie}
\end{figure}

Fig.~\ref{curie} compares the magnetic susceptibility of the large $d(Fe-N)$ and short $d(Fe-N)$  structures over a wide temperature range. The very small value and weak temperature dependence of $\chi$ in the short $d(Fe-N)$ structure demonstrates that this structure has a low-spin ground state, while the Curie-like temperature dependence of the susceptibility in the long $d(Fe-N)$ structure clearly shows that the local moment is well formed, even at the highest temperature studied   here.  

\begin{figure}[b]
\includegraphics[width=\linewidth,trim={3.5cm 1.5cm 1.5cm 2.5cm},clip]{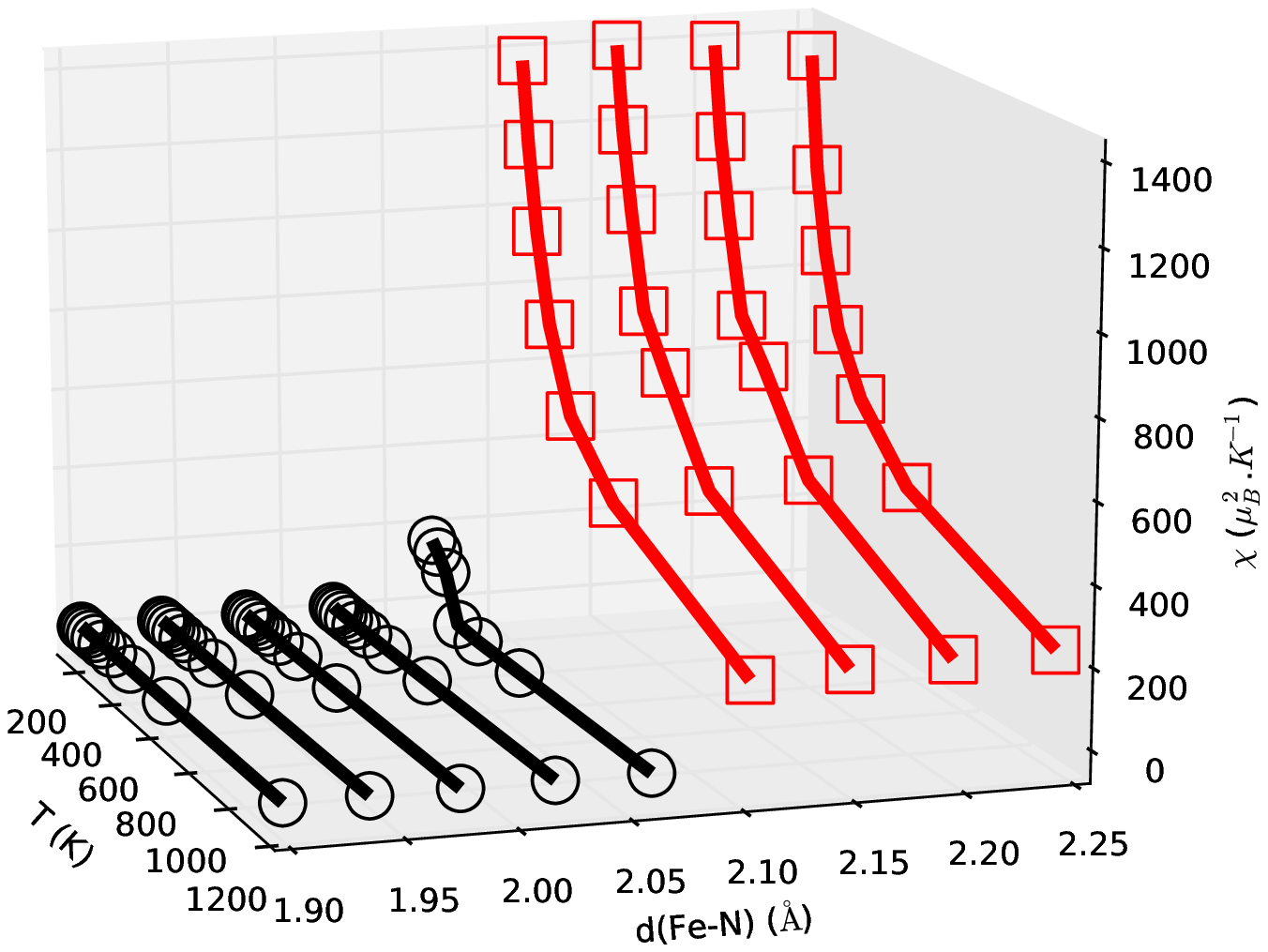}
\caption{Dependence of magnetic susceptibility on temperatures and average Fe-to-N bond length.} \label{control}
\end{figure}

Fig.~\ref{control} shows the temperature dependence of the magnetic susceptibility for a series of structures with different average Fe-to-N bond lengths. This figure demonstrates the existence of a  critical value of the mean bond length $\approx 2.10\AA$ above which the molecule is in the HS state  and below which it is in the LS state. 

\begin{figure}
\includegraphics[width=\linewidth,clip]{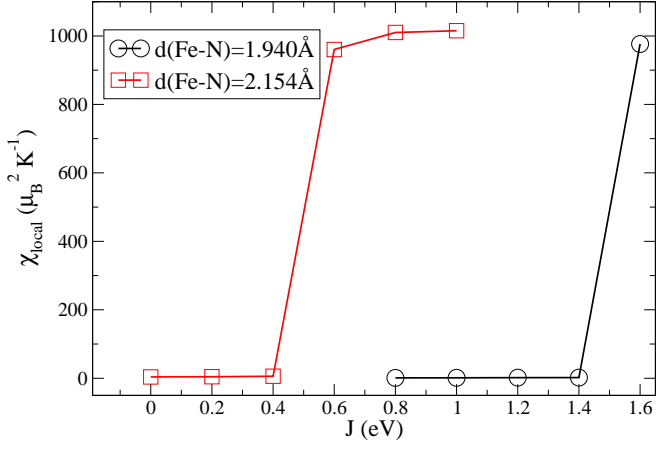}
\caption{Magnetic susceptibility as function of exchange J from DFT+DMFT calculations with $U=5.0eV$ and $T=387K$.} \label{J}
\end{figure}

Fig.~\ref{J} shows the magnetic susceptibility calculated as a function of Hunds coupling $J$  at a temperature $T=387K$ slightly higher than room temperature for the short $d(Fe-N)$ and long $d(Fe-N)$ structures. We see that for very small $J$, the ground state of  both the long-bond and the short-bond structure is LS. For both structures,  as $J$ is increased a LS-HS transition occurs. The critical value of $J$ is $\approx 0.4eV$ for the long $d(Fe-N)$ structure; this value is smaller than the physically reasonable $J\sim 0.9eV$ expected for $Fe$. On the other hand for the short $d(Fe-N)$ structure an unphysically large $J\approx 1.4eV$ is required to drive the transition.  Thus, for a large window of J (0.5 eV to 1.5 eV) including  physically reasonable values, DFT+DMFT is able to separate two spin states based on atomic structure.

\section{Fe site electronic configuration}
\begin{figure}
\includegraphics[width= \linewidth,clip]{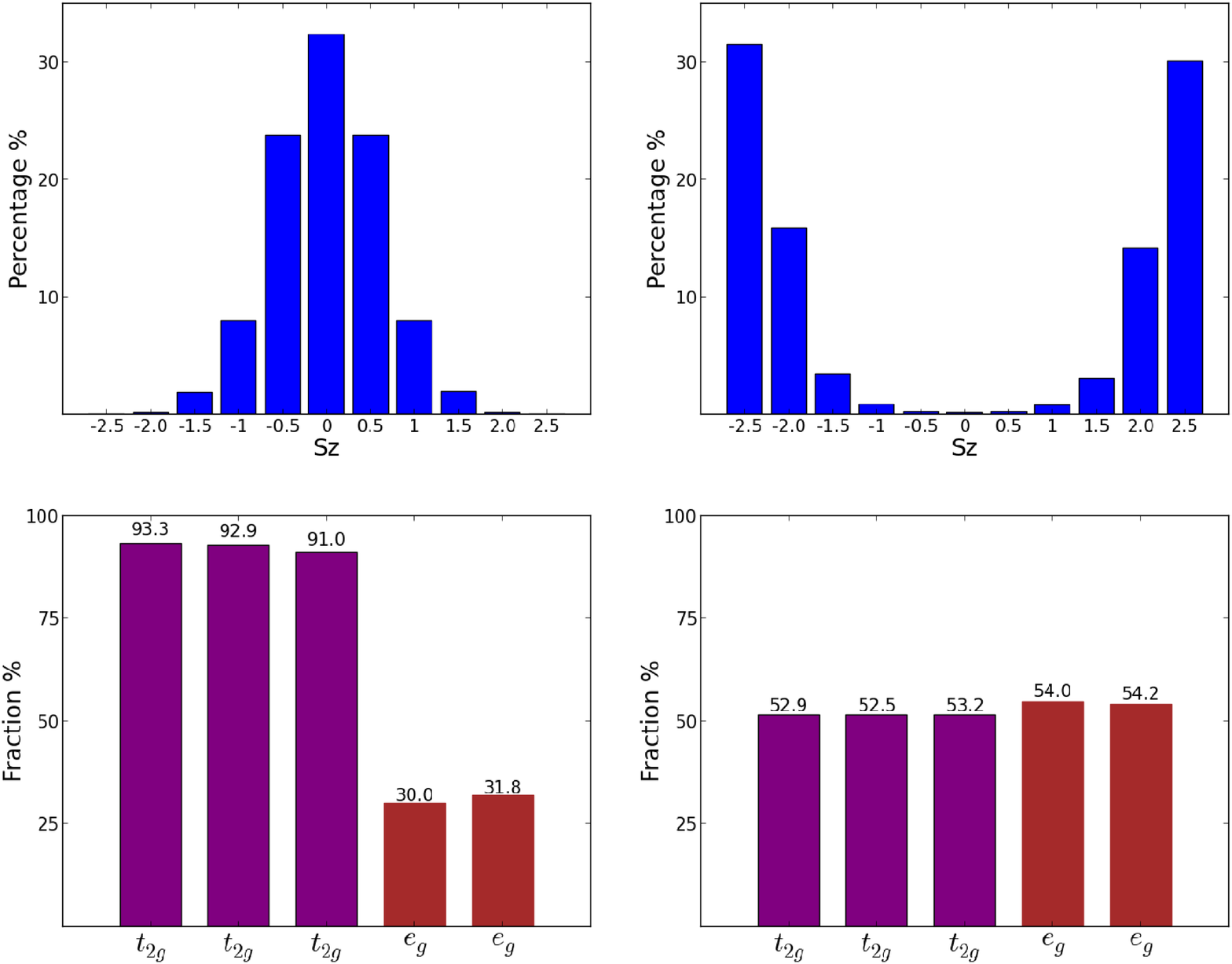}
\caption{Spin (upper left and upper right) and (lower left and lower right) sector statistics for LS ((a) and (c)) and  HS ((b) and (d))  from DFT+DMFT calculations at $U=5.0eV$, $J=0.85eV$ and $T=387K$.} \label{his_occ}
\end{figure}

Aspects of the Fe site electronic configuration can be directly measured in the DMFT calculations \cite{Haule07,Gull11_review}. A measurements of the fraction of imaginary time that the impurity spends in a given configuration in the simulation yields the fractional weight of  the configuration in the density matrix. Panels (a) and (b) of  Fig.~\ref{his_occ} show the occupancies of different $S_z$ states. In the LS, $S_z=0$ is the most probable configuration, while for HS the most probable state is  $S_z=\pm2.5$. Orbital occupations (panels (c) and (d) in Fig.~\ref{his_occ}) also show the distinction between two states. In LS, t$_{2g}$ orbitals are close to full occupation and e$_g$ orbitals are less than half occupied; but in HS, all orbitals are close to half occupation. 

\begin{figure}
\includegraphics[width= \linewidth,clip]{S4.eps}
\caption{Crystal field splitting as function of Fe-to-N distance.} \label{cry}
\end{figure}

\section{Mean d occupancy and double counting}
\begin{figure}
\includegraphics[width= \linewidth,clip]{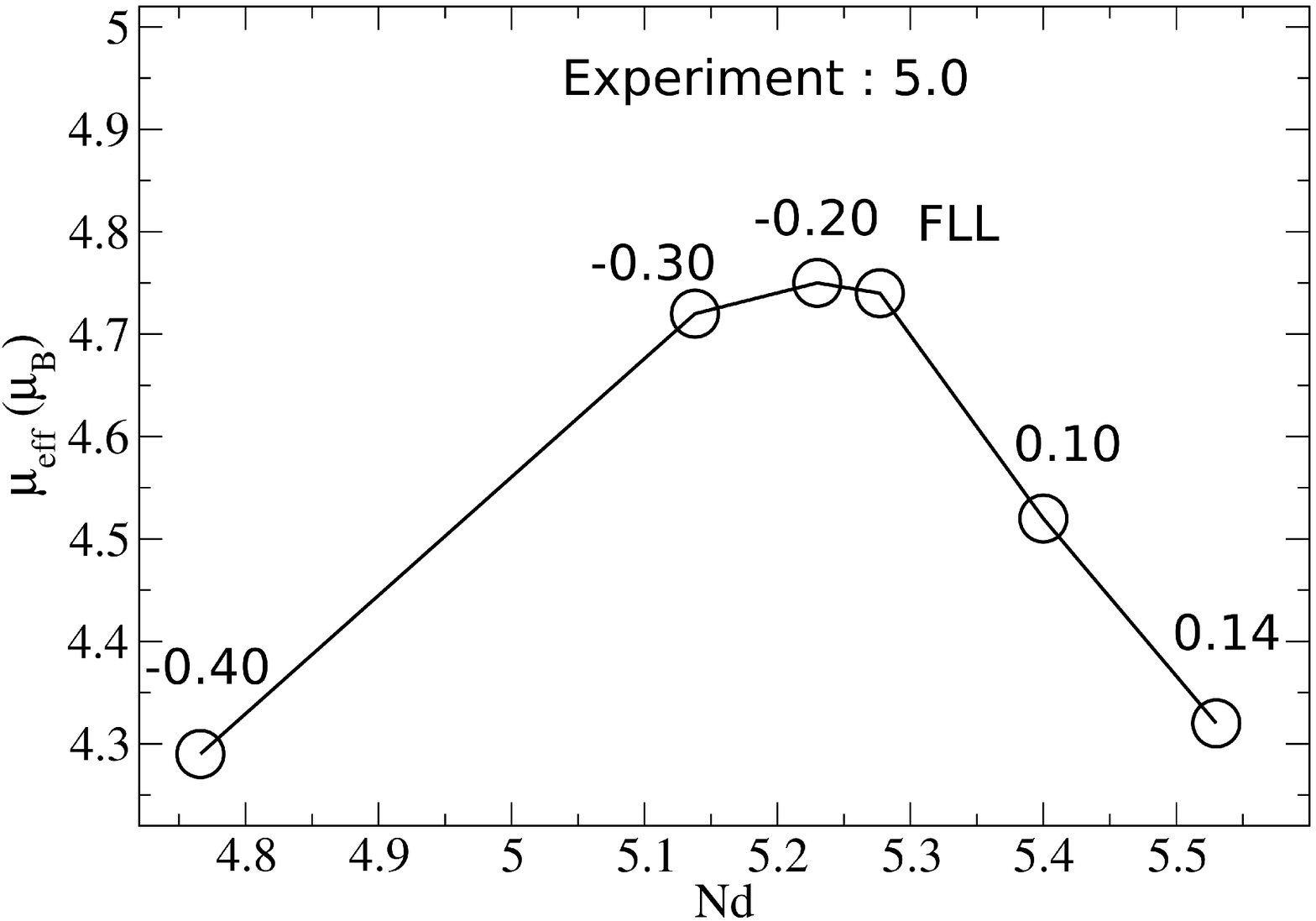}
\caption{Dependence of effective magnetic moment $\mu_{eff}=\sqrt{\chi T}$ obtained from  on occupancy $N_d$ of Fe d orbitals in high spin state at $U=5.0eV$, $J=0.85eV$ and $T=387K$. Different values of $N_d$ are obtained by varying the coefficient  $U^\prime$ in the double counting formula of the main text.} \label{muvsNd}
\end{figure}

\begin{figure*}
\includegraphics[width=\linewidth]{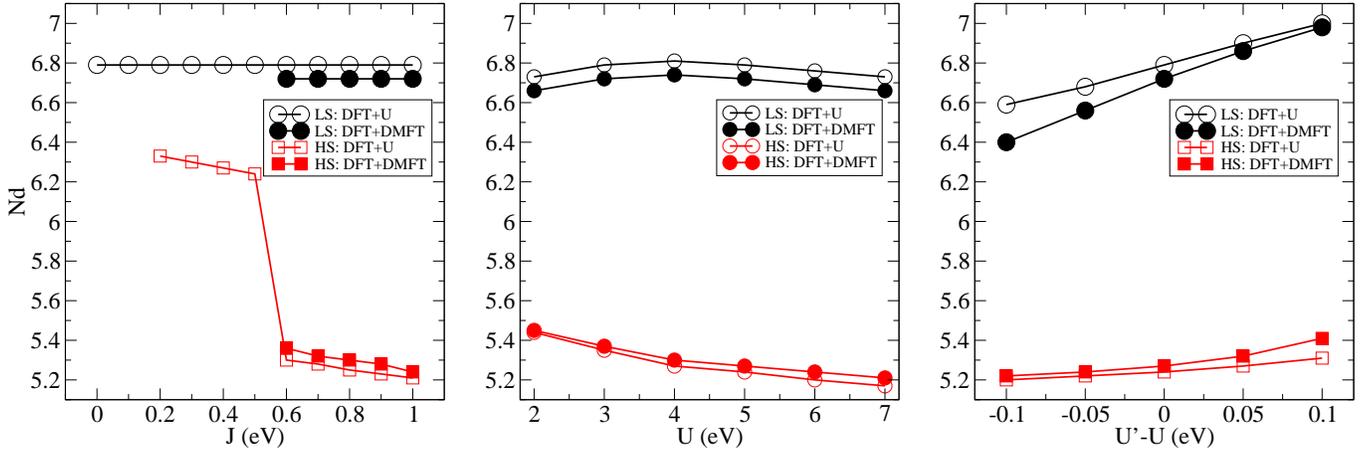}
\caption{N$_d$ from DFT+U and DFT+DMFT calculations performed for short $d(Fe-N)$ (calculation restricted to LS state) and long $d(Fe-N)$ (calculation restricted to HS state) structures, with various U, J and U'. Temperature is 387K in DMFT calculations.} \label{Nd}
\end{figure*}

\begin{figure}
\includegraphics[width= 0.85\linewidth,clip]{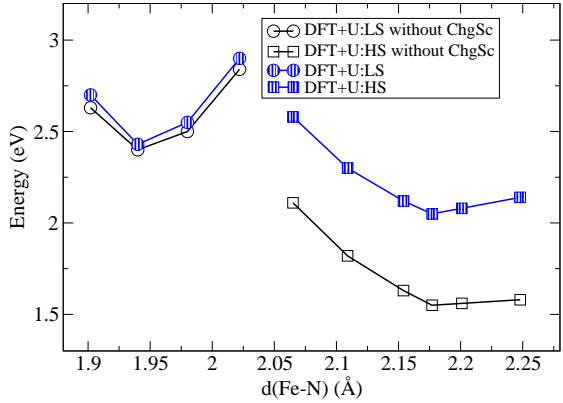}
\caption{Total energy of LS and HS from DFT+U calculations with and without charge self-consistency. U and J are fixed at 5.0 eV and 0.85 eV.} \label{ChgScf}
\end{figure}

The LS and HS states differ in relative occupancies of the $t_{2g}$ and $e_g$ orbitals. This difference is the result of a competition between Hunds exchange J and crystal (ligand) field splitting, which arises mainly from hybridization between Fe $d$ and $N$-$p$ orbitals. The crystal field splittings were calculated as the difference between average energies of t$_{2g}$ and e$_g$ orbitals obtained from the on-site terms in the projection of the DFT Hamiltonian onto our MLWF basis. The calculated crystal field splittings are shown in Fig.~\ref{cry} and are clearly seen to be strongly influenced by the mean Fe-to-N distance

Previous work \cite{Wang12,Dang14a,Dang14b} has shown that the occupancy $N_d$ of the correlated orbitals provides a useful characterization of the beyond-DFT many-body state, and can be varied by changing the double counting coefficient $U^\prime$ \citep{Park2014}.   $N_d$ is also related to the effective magnetic moment of the HS state and thus can be inferred from experiment.  We have computed the effective magnetic moment $\mu_{eff}=\sqrt{\chi T}$ by performing fully charge self consistent DFT+DMFT calculations using different values of  $U^\prime$.  Measurements in the HS state indicate  $\mu_{eff}=5\mu_B$ \cite{Shi2009}. As we can see from Fig.~\ref{muvsNd}, for all Nd we studied in the HS state (4.7$\sim$5.5), our calculated effective magnetic moments are  smaller than the experimental value.  The standard FLL double counting $U^\prime=U$ yields an $N_d\approx 5.25$  and  an effective magnetic moment close to the maximum value and  to experiment.

We find that the d-occupancy is much higher in the LS state than in the HS state. The physics is that in the HS state the longer  $d(Fe-N)$ distance weakens the hybridization while the  minority spin orbitals are pushed to a very high energy and thus hybridize much less with the ligands. Fig.~\ref{Nd} shows $N_d$ computed with Hartree-Fock (DFT+U) for the short $d(Fe-N)$ (``LS'') and long $d(Fe-N)$ (``HS'') structures. Note that  for the long $d(Fe-N)$ structure  the HS state is at least metastable within the Hartree-Fock approximation at all $J$ which were studied, unlike the DMFT calculation where  a HS-to-LS transition occurs as J is reduced below $\approx 0.4eV$ in the long $d(Fe-N)$ structure.  This is why the DMFT curves only extend down to $J=0.6$ in the left panel.

We see  that where the HS state  is stable in both methods, DFT+U and DFT+DMFT give reasonably consistent pictures for the value of $N_d$ in the two structures, and that $N_d$ is much more sensitive to the structure than it is to the interaction parameters. Of particular interest is the $J$ dependence of $N_d$ in the structure with the long $d(Fe-N)$. In this structure, within the Hartree-Fock approximation a level crossing between two HS states with different $N_d$  occurs at a $J\approx 0.5eV$. A simplified picture of this level crossing is that as J is decreased  it becomes favorable to add one electron to a minority-spin $t_{2g}$ level, thus increasing $N_d$ and reducing the moment by $\approx \frac{1}{2}$. We emphasize that this metastable high $N_d$, high spin state is found in the HF calculations but not in the DMFT calculations. A truly low-spin state with $N_d \gtrsim 6$ is also metastable within the Hartree-Fock approximation, but has higher energy than the state shown here.    Level crossings in d-occupancy related to different spin states were also reported in a recent study  of the heme molecule. \cite{Weber2013}

\section{The effect of charge self-consistency}
Charge self-consistency can have a substantial impact on the total energy if there is a major rearrangement of charge from the initial density which was used to build the Kohn-Sham potential. The starting point of our DFT+DMFT calculation is a spin-unpolarized DFT calculation, which is inherently low spin. While this density will be very similar to the converged LS solution, it will be quite different from the converged HS solution.  This is due both to the change in orbital occupancy  and the overall large change of $N_d$ across the LS-HS transition.  In Fig. ~\ref{ChgScf} we illustrate this effect by comparing the total energy within DFT+U both with and without charge self-consistency. As expected, the changes are relatively small for the LS state and relatively large for the high spin state.

\bibliography{Collection}